\documentclass[doublecol,linenumbers]{epl2}
\usepackage{graphicx}
\usepackage[english]{babel}
\usepackage{latexsym,amsmath,amscd,amssymb,graphicx,amsbsy,color,xcolor,amsfonts,amsthm}

\title{Relativistic  Newtonian Dynamics for Objects and Particles}
\shorttitle{Relativistic  Newtonian Dynamics }
\author{ Y. Friedman}
\institute{
   Jerusalem College of Technology\\Jerusalem, Israel}


\abstract{
 Relativistic Newtonian Dynamics (RND) was introduced in a series of recent papers by the author, in partial cooperation with J. M. Steiner.  RND was capable of describing non-classical behavior of motion under a central attracting force. {RND incorporates the influence of potential energy on spacetime in Newtonian dynamics, treating gravity as a force in flat  spacetime.} It was shown that this dynamics predicts accurately gravitational time dilation, the anomalous precession of Mercury and  the periastron advance of any binary.
 In this paper the model is further refined and extended to describe also the  motion of both objects with non-zero mass and massless particles, under a conservative attracting force. It is shown that for any conservative force a properly defined energy is conserved on the trajectories and if this force is central, the angular momentum is also preserved. An RND equation of motion is derived for motion under a conservative force. As an application, it is shown that RND predicts accurately also the Shapiro time delay - the fourth test of GR.}

\begin{document}
\maketitle

\section{Introduction}

A new relativistic model  incorporating the influence of potential energy on spacetime in  Newtonian dynamics for  motion of non-zero mass objects under a central force, named \textit{Relativistic Newtonian Dynamics} (RND), was introduced recently \cite{FSMerc}, \cite{Fcent} and \cite{FSBin}. In these papers it was shown that this dynamics was tested successfully to  predict accurately the gravitational time dilation, the anomalous precession of Mercury and the periastron advance of any binary.

 All the above tests of RND, were connected solely with the \textit{trajectories} {(not with the time dependence of the position on the trajectory)} of \textit{massive} objects moving under a static conservative attracting \textit{central} force vanishing at infinity. {Unlike in the above tests, the time delay caused by the slowing passage of light as it moves over a finite distance through a spatially changing gravitational potential (Shapiro time delay), must consider the time dependence of the position on the trajectory.}

  In this paper the RND model is extended to derive the equation of motion under a \textit{conservative} attracting force for both objects with non-zero mass and \textit{massless} particles. The energy conservation equation for any conservative attractive force and the angular momentum conservation for such a central force, are derived. In its present form, the model is capable to describe accurately the position on the trajectory of the object/particle at any given time. This extended model is then applied  to obtain the correct Shapiro time delay, the fourth test of GR.

\section{RND Energy  under a Conservative Force}

 Consider the motion under a conservative force with a negative potential $U(\mathbf{x})$ which is time independent in an inertial lab frame $K$ and vanishing at infinity. For example, $U(\mathbf{x})$ may be the Newtonian potential of a gravitational field generated by several planets or stars, in an inertial frame $K$ {with its origin at the center of mass} of this assembly. Introduce a vector field $\mathbf{n}(\mathbf{x})$ which at any space point $\mathbf{x}$ is the  normalized vector in the direction of the gradient $\nabla U(\mathbf{x})$, i.e.
\begin{equation}\label{direction_force}
 \mathbf{n}(\mathbf{x})=\nabla U(\mathbf{x})/|\nabla U(\mathbf{x})|.
\end{equation}

As suggested in \cite{Fcent}, {using an extended version of the Equivalence Principle, the influence due to the potential energy on local spacetime in the neighbourhood of some point $\mathbf{x}$ can be quantified by the relativistic space contraction and time dilation}  due to the  velocity (kinetic energy) of the escape frame at $\mathbf{x}$. Using this fact, it was shown {there} that the potential energy influences time intervals and space increments in the direction of $\mathbf{n}(\mathbf{x})$, while the space increments transverse to $\mathbf{n}(\mathbf{x})$ are not influenced by this potential energy.

By analogy to the principle of least action, a variational principle that defines the path of motion as the path with the least value of some action, the motion in such an influenced spacetime can be viewed  as the motion along a geodesic (path with least distance) with respect to some metric. By the above observation this metric is of the form
\begin{equation}\label{metric2}
  ds^2=f(\mathbf{x})(cdt)^2-g(\mathbf{x}) d\mathbf{x}_n^2-d\mathbf{x}_{tr}^2,
\end{equation}
where the two vector projections of the space increment in the parallel and transverse directions to $\mathbf{n}(\mathbf{x})$ are
\begin{equation}\label{parralel}
   d\mathbf{x}_n=(d\mathbf{x}\circ\mathbf{n})\mathbf{n}, \;\;d\mathbf{x}_{tr}=d\mathbf{x}-(d\mathbf{x}\circ\mathbf{n})\mathbf{n}
 \end{equation}
respectively, with $\circ$ denoting the Euclidean dot product in $R^3$  and the positive valued functions $f(\mathbf{x}),g(\mathbf{x})$ describe the influence of the potential energy on  time intervals and space increments in the $\mathbf{n}(\mathbf{x})$ direction.

Consider first the time dilation due to the influence of potential energy of a clock resting at some point $\mathbf{x}$ in $K$. Since for such clock $d\mathbf{x}=0$, from (\ref{metric2})  $ds=\sqrt{f(\mathbf{x})}cdt$ implying that the time dilatation of such a clock is $c\frac{dt}{ds}=\frac{1}{\sqrt{f(\mathbf{x})}}$. Assume that this time dilation is the same to that of a clock resting in the escape frame \cite{Fcent}. This time dilation is defined by the gamma-factor $\tilde{\gamma}$ of the escape velocity $\mathbf{v}_e$,  where
\begin{equation}\label{gammatil}
  \tilde{\gamma}=\frac{1}{\sqrt{1-v_e^2/c^2}}=\frac{1}{\sqrt{1-u}},\;\;\; u(\mathbf{x})=\frac{-2U(\mathbf{x})}{mc^2}
\end{equation}
and $u(\mathbf{x})$ is the negative of the \textit{dimensionless potential energy}. Thus,
\begin{equation}\label{fxdef}
  f(\mathbf{x})=1-u(\mathbf{x}).
\end{equation}
This formula {is true according the model for any gravitational field and in particular it reduces to the (experimentally tested)} gravitational time dilation in the vicinity of a non-rotating massive spherically symmetric object, as derived from the Schwarzschild metric.

The motion of an object of mass $m>0$  along a time-like geodesic $\mathbf{X}(\lambda)=(ct(\lambda),\mathbf{x}(\lambda))$ in spacetime is parameterized by an \textit{affine parameter} $\lambda$ on the trajectory, chosen to be the arc length $s$ defined by (\ref{metric2}). For massless objects (photons) moving along light-like geodesics $ds=0$,  this affine parameter is redefined as in \cite{MTW} p.575.

Since the metric (\ref{metric2}) is independent of $t$, the vector $\mathbf{K}_0=(1,0,0,0)$ is a Killing vector (\cite{MTW}, p.651) and the scalar product $\dot{\mathbf{X}}\cdot\mathbf{K}_0$ with respect to this metric is conserved on the trajectory. Thus, on any trajectory there exists a constant $k$ related to the total energy on it such that
\begin{equation}\label{dtdsC}
c f(\mathbf{x}) \dot{t}=k,\;\;\; c\dot{t}=\frac{k}{1-u(\mathbf{x})},
\end{equation}
where $dot$ denotes  differentiation  with respect to $\lambda$.

The four-velocity on the trajectory is $\dot{\mathbf{X}}=\frac{d\mathbf{X}}{d\lambda}=(c\dot{t},\dot{\mathbf{x}})$. Using  (\ref{metric2}),(\ref{fxdef}) and (\ref{dtdsC}) its norm is
 \begin{equation}\label{4velNorm2}
  \dot{\mathbf{X}}^2=\frac{k^2}{1-u(\mathbf{x})}-g(\mathbf{x})\dot{\mathbf{x}}_n^2-\dot{\mathbf{x}}_{tr}^2=\epsilon,
\end{equation}
where $\epsilon= 1$ for objects with non-zero mass and $\epsilon= 0$ for massless particles.
For objects with non-zero mass multiplying equation (\ref{4velNorm2}) by $(1-u(\mathbf{x}))$  one obtains
\begin{equation}\label{RNDenergy}
 (1-u(\mathbf{x}))g(\mathbf{x})\dot{\mathbf{x}}_n^2+(1-u(\mathbf{x}))\dot{\mathbf{x}}_{tr}^2-u(\mathbf{x})=k^2-1\,.
\end{equation}

This formula can be considered as the RND  dimensionless energy conservation equation on the trajectory. The first two terms of the left side are the relativistically corrected  dimensionless kinetic energy and the third term is  the dimensionless potential energy. The right side is the dimensionless total energy $\mathcal{E}=2E/mc^2=k^2-1$.

Introduce a function $\Phi$ mapping velocities $\mathbf{v}=\frac{d\mathbf{x}}{dt}$ at some point $\mathbf{x}$ in the spacetime influenced by the potential energy to the corresponding velocity in the lab frame. As  shown in \cite{Fcent}, $\Phi$ {acts} on the  velocity components  as $\Phi(\mathbf{v}_n)=\mathbf{v}_n$ and $\Phi(\mathbf{v}_{tr})=\sqrt{1-u(\mathbf{x})}\mathbf{v}_{tr}$. The relativistically corrected  dimensionless kinetic energy (the first two terms in (\ref{RNDenergy})) is $\Phi(\dot{\mathbf{x}})^2$. Since $\dot{\mathbf{x}}=\dot{t}\mathbf{v}$ we obtain
\begin{equation*}
  \Phi(\dot{\mathbf{x}})^2=\dot{t}^2\Phi(\mathbf{v})^2=\dot{t}^2\mathbf{v}_n^2 +\dot{t}^2(1-u(\mathbf{x}))\mathbf{v}_{tr}^2\end{equation*}
\begin{equation*}=\dot{\mathbf{x}}_n^2 +(1-u(\mathbf{x}))\dot{\mathbf{x}}_{tr}^2\,.
\end{equation*}

Comparing this to (\ref{RNDenergy}) one obtains
\begin{equation}\label{metrCoef}
 g(\mathbf{x})=\frac{1}{1-u(\mathbf{x})},
\end{equation}
hence, the RND dimensionless energy conservation equation for an object with non-zero mass becomes
\begin{equation}\label{RNDenergyM}
 \dot{\mathbf{x}}_n^2+(1-u(\mathbf{x}))\dot{\mathbf{x}}_{tr}^2-u(\mathbf{x})=\mathcal{E}.
\end{equation}

For a massless particle, the analysis is only applicable when restricted to a gravitational potential. For such a potential one can define a \textit{reduced potential} or potential per unit mass as
 \begin{equation}\label{reduredPot}
   \hat{U}=\frac{U}{m}.
\end{equation}
For such {a} potential, the dimensionless potential defined  by (\ref{gammatil})  is $u(\mathbf{x})=-2\hat{U}/c^2$ is mass independent.

 For a massless particle ($\epsilon=0$), multiplying (\ref{4velNorm2}) by $(1-u(\mathbf{x}))$ and using (\ref{metrCoef}) one obtains
 \begin{equation}\label{RND decompNGPh}
 \dot{\mathbf{x}}_n^2+(1-u(\mathbf{x}))\dot{\mathbf{x}}_{tr}^2=k^2.
\end{equation}
 Even though gravitation does not act directly on the photon as a force, since its mass is zero, which is expressed in the missing term $-u(\mathbf{x})$ of (\ref{RNDenergyM}), its momentum and angular momentum are not zero and its motion is affected by the influence of the gravitational potential on spacetime expressed by the relativistic term  in (\ref{RND decompNGPh}).

Thus, the RND dimensionless energy conservation  equation  for both objects with non-zero mass and massless particles is
\begin{equation}\label{RND decompGen2}
 \dot{\mathbf{x}}_n^2+(1-u(\mathbf{x}))(\dot{\mathbf{x}}_{tr}^2+\epsilon )=k^2.
\end{equation}
In RND the reduced energy (obtained by multiplying the dimensionless energy by $c^2/2$)
 \begin{equation}\label{RNDHamit}
   H(\mathbf{x},\dot{\mathbf{x}})=\frac{c^2\dot{\mathbf{x}}^2}{2}+\hat{U}(\mathbf{x})(\dot{\mathbf{x}}^2-(\dot{\mathbf{x}}\circ\mathbf{n})^2)+\epsilon \hat{U}(\mathbf{x})
 \end{equation}
is conserved on the trajectory. Note that the relativistic addition to a classical reduced total energy, expressed in the middle term  on the right-hand side, depends both on the potential at that point and the velocity of the moving object.

\section{RND Equation of Motion under a Conservative Force}

By use of (\ref{parralel})  equation (\ref{RND decompGen2}) can be rewritten as
\begin{equation}\label{RND decompGen3}
 \dot{\mathbf{x}}^2-\epsilon u(\mathbf{x})-u(\mathbf{x})(\dot{\mathbf{x}}^2-(\dot{\mathbf{x}}\circ\mathbf{n})^2)-\epsilon =k^2.
\end{equation}

The equation of motion is obtained by differentiating (\ref{RND decompGen3}) by  $\lambda$ which leads to
\begin{equation*}2\ddot{\mathbf{x}}\circ\dot{\mathbf{x}}- \epsilon\dot{u}-\dot{u}\dot{\mathbf{x}}_{tr}^2-2u(\ddot{\mathbf{x}}\circ\dot{\mathbf{x}}-(\dot{\mathbf{x}}\circ\mathbf{n})(\ddot{\mathbf{x}}\circ\mathbf{n}+\dot{\mathbf{x}}\circ\dot{\mathbf{n}}))=0,\end{equation*}
Substituting $u(\mathbf{x})=-2\hat{U}/c^2$ and its derivative $\dot{u}=-\frac{2\nabla \hat{U}\circ\dot{\mathbf{x}}}{c^2}$, one obtains
\begin{equation}\label{RND1}
 \ddot{\mathbf{x}}+\frac{2\hat{U}}{c^2}\ddot{\mathbf{x}}_{tr}=-\frac{\epsilon\nabla \hat{U}}{c^2}-\frac{\nabla \hat{U}\dot{\mathbf{x}}^2_{tr}}{c^2}+
 \frac{2\hat{U}}{c^2}(\dot{\mathbf{x}}\circ\dot{\mathbf{n}})\mathbf{n}.
\end{equation}

Decomposing this vector equation  into the parallel and  perpendicular components to $\mathbf{n}$, the  perpendicular component is
\begin{equation}\label{PerpRNDeqn}
 \left(1+\frac{2\hat{U}(\mathbf{x})}{ c^2}\right)\ddot{\mathbf{x}}_{tr}=0.
\end{equation}
implying that $\ddot{\mathbf{x}}_{tr}=0$ for $\mathbf{x}$ outside the Schwarzschild horizon defined as the surface where the escape velocity is $c$.

 {This implies that the acceleration  with $\mathbf{x}$ twice differentiated with respect to the affine parameter $\lambda$ is in the direction of the force $\mathbf{n}$. Indeed, this property is specific only for the affine parameters, as follows. Let $\tilde{\lambda}=f(\lambda)$ be any other parameter on the trajectory. Then $\frac{d\mathbf{x}}{d\lambda}=f'(\lambda)\frac{d\mathbf{x}}{d\tilde{\lambda}}$ and $\frac{d^2\mathbf{x}}{d\lambda^2}=f"(\lambda)\frac{d\mathbf{x}}{d\tilde{\lambda}}+f'(\lambda)^2\frac{d^2\mathbf{x}}{d\tilde{\lambda^2}}$, implying that $\frac{d^2\mathbf{x}}{d\tilde{\lambda^2}}$ is in the direction of $\mathbf{n}$ only if $f"(\lambda)\equiv 0$. This condition implies \cite{MTW} that $\tilde{\lambda}$ is an affine parameter. }

Thus, for motion outside the Schwarzschild horizon  the \textit{RND equation of motion} for any object/particle is
\begin{equation}\label{RND1finals}
 c^2\ddot{\mathbf{x}}=-\epsilon\nabla \hat{U}-\dot{\mathbf{x}}^2_{tr}\nabla \hat{U}+ 2\hat{U}(\mathbf{x})(\dot{\mathbf{x}}\circ\dot{\mathbf{n}})\mathbf{n}.
\end{equation}
Noting that $\frac{d}{d\lambda}=\frac{k}{c(1-u)}\frac{d}{dt}$ from (\ref{dtdsC}) and remembering that $\mathbf{v}=\frac{d\mathbf{x}}{dt}$ one obtains the RND equation of motion in the space and time with respect to the lab frame $K$ as
\begin{equation*}
 \frac{k^2}{1-u}\frac{d}{dt}\left(\frac{\mathbf{v}}{1-u}\right)=\end{equation*}
 \begin{equation}\label{RNDf}=- \epsilon\nabla \hat{U}-\frac{k^2\mathbf{v}^2_{tr}\nabla \hat{U} }{c^2(1-u)^2}+
 \frac{2k^2\hat{U}}{c^2(1-u)^2}(\mathbf{v}\circ\frac{d\mathbf{n}}{dt})\mathbf{n}.\end{equation}
 Note that the acceleration $\frac{d^2\mathbf{x}}{dt^2}$ is not in the direction of $\mathbf{n}$, {as expected, since as implied by (\ref{dtdsC}) time $t$ is not an affine parameter.}

 The first relativistic correction depends on $-\nabla \hat{U}$- the classical acceleration caused by the force, and the second one depends on the reduced potential $\hat{U}$. Unlike in the classical Newtonian equation of motion involving only the force $\mathbf{F}$, the RND equation involves both the potential $U$ (as in Schr\"{o}dinger quantum dynamics equation) as well as  $\nabla U$.

\section{ RND under a Central Force}

Let $U(r)$ be a potential of a central force which is attractive and static in an inertial lab frame $K$ with the origin at the center of the force. Assume that $U(r)$ is negative and vanishes at infinity. As shown in \cite{Fcent},  in this case the time intervals and the radial space increments are influenced by the potential  energy, while the space increments transverse to the radial direction are not affected.

Introduce spherical coordinates $r,\varphi,\theta$ in $K$. In this case $\mathbf{n}$ defined by (\ref{direction_force}) is the radial direction and thus $\dot{\mathbf{x}}_n=\dot{r}$. Using the results of Section 2, the influence of the potential energy on spacetime in the neighbourhood of any  point is described  by the metric
\begin{equation*}
  ds^2=(1-u(r))(cdt)^2-\frac{1}{1-u(r)} (dr)^2-
\end{equation*}
\begin{equation}\label{metric}
  -r^2((d\theta)^2+\sin^2 \theta (d\varphi)^2).
\end{equation}
As above, assume that the motion of an object/particle is along the geodesic in spacetime $\mathbf{X}(\lambda)=(t(\lambda),r(\lambda),\varphi(\lambda),\theta(\lambda))$ parameterised by the affine parameter $\lambda$. Then the  four-velocity is $\dot{\mathbf{X}}=\frac{d\mathbf{X}}{d\lambda}=(c\dot{t},\dot{r},\dot{\varphi},\dot{\theta})$.
It can be shown \cite{Gron}p.172,3  that if initially the position and velocity of the object/particle are in the plane $\theta=\pi/2$, they will remain in this plane during the motion. Thus, one may chose the coordinate system so that that $\theta=\pi/2,\;\dot{\theta}=0$.

Using that the metric is independent of $\varphi$, the vector $\mathbf{K}_3=(0,0,1,0)$ is a Killing vector
implying that
\begin{equation}\label{Jdef}
  J=r^2\dot{\varphi}
\end{equation}
remains constant on the trajectory, where $cJ$ has meaning of angular momentum per unit mass for objects with non-zero mass and $J$ has units of length.

This implies that $\dot{\mathbf{x}}^2_{tr}=\frac{J^2}{r^2}$ and
one can rewrite the RND dimensionless energy conservation  equation (\ref{RND decompGen2}) as
\begin{equation}\label{RND decompGen}
 \dot{r}^2 +(1-u(r))\left(\frac{J^2}{r^2}-\epsilon \right)=k^2.
\end{equation}
Using (\ref{Jdef}) and (\ref{RND decompGen}) one obtains the parameter-free equation for the trajectory  $r(\varphi)$
\begin{equation}\label{drdphi}
  \left(\frac{J}{r^2}\frac{dr}{d\varphi}\right)^2+(1-u(r))\left(\frac{J^2}{r^2}-\epsilon \right)=k^2.
\end{equation}
This equation with $\epsilon=1$ was used in the previous papers \cite{FSMerc}, \cite{Fcent} and \cite{FSBin} for objects with non-zero mass.

\section{Shapiro Time Delay}

The Shapiro time delay (or gravitational time delay),  the fourth  test of GR, describes the slowing of light as it moves over a finite distance through a change in the gravitational potential of a massive object $M$.

Consider  the motion of  a massless photon ($\epsilon=0$) from a point $A$ to a point $B$ in the gravitational field of a spherically symmetric massive  object of mass $M$.
Denote by $r_0$ the distance from the point $P$ on the trajectory closest to the massive object (Sun), (see Figure 1).

\begin{figure}[h!]
\centering
 \scalebox{0.28}{\includegraphics{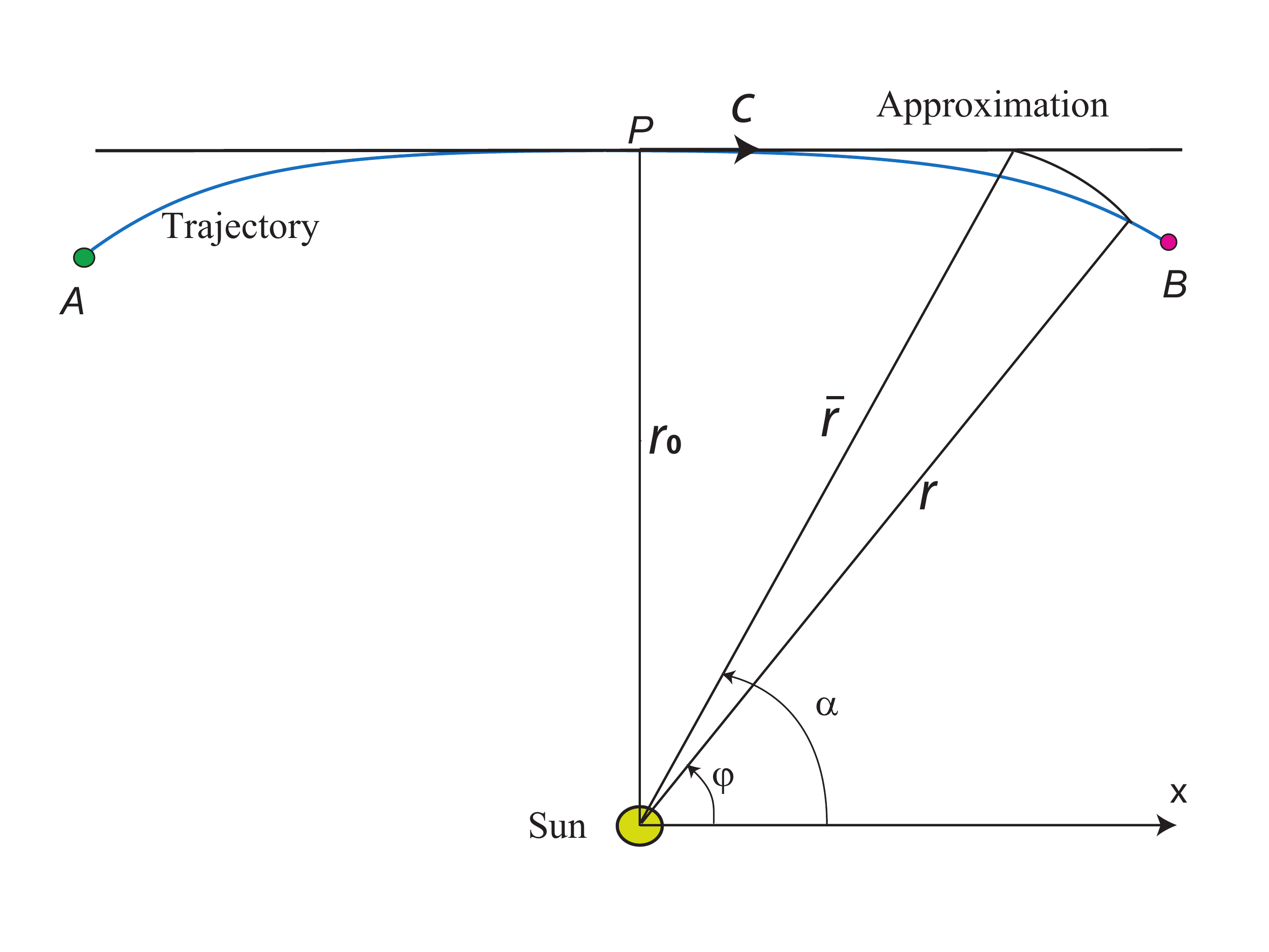}}
 \caption{The bending trajectory a massless particle, the straight line approximation and associated angles $\varphi,\alpha$ }\label{Prec}
\end{figure}

  By introducing the Schwarzschild radius $r_s= \frac{2GM}{c^2}$, the dimensionless potential energy, defined by (\ref{gammatil}) is
\begin{equation}\label{schwatz}
u(r)=\frac{r_s}{r}.
\end{equation}
Since $\frac{dr}{d\varphi}=0$ at the point $P$, from (\ref{drdphi}) the integrals of motion relation $J/k$ (independent of the parametrization) is
\begin{equation}\label{Impact}
  \frac{J}{k}=-\frac{r_0}{\sqrt{1-r_s /r_0}}=-b,
\end{equation}
where $b$ is called the \textit{impact parameter} of the trajectory.

Using this and equations (\ref{dtdsC})and (\ref{Jdef})  one obtains
\begin{equation}\label{dphidt}
  \frac{d\varphi}{dt}=\frac{\dot{\varphi}}{\dot{t}}=-\left(1-\frac{r_s}{r}\right)\frac{cb}{r^2},
\end{equation}
implying that
\begin{equation*}
  cdt=\frac{-r^2}{(1-r_s/r)b}d\varphi \,.
\end{equation*}

The time of passage from the point $P$ to $B$ is
\begin{equation*}
  c(T_B-T_P)=\int_{\varphi_B}^{\pi/2}\frac{r^2}{(1-r_s/r)b}d\varphi \,.
\end{equation*}
For any angle $\varphi$ on the trajectory one may associate an angle $\alpha(\varphi)$ for which $r(\varphi)=\bar{r}(\alpha)$ where $\bar{r}(\alpha)=\frac{r_0}{\sin \alpha}$ is the straight line  approximation of the trajectory at the point $P$ (see Figure 1).
This suggests the substitution $r=\frac{r_0}{\sin\alpha}$
which implies
\begin{equation*}
  c(T_B-T_P)=\frac{r_0^2 }{b}\int_{\alpha_B}^{\pi/2}\frac{1}{\sin^2\alpha(1-\frac{r_s}{r_0}\sin\alpha)}d\alpha =
\end{equation*}
\begin{equation*}=\frac{r_0^2 }{b}\int_{\alpha_B}^{\pi/2}\left(\frac{1}{\sin^2\alpha}+\frac{r_s/r_0}{\sin\alpha}+\frac{r_s^2 /r_0^2}{(1-\frac{r_s}{r_0}\sin\alpha)}\right)d\alpha=\end{equation*}
\begin{equation*}=\frac{r_0^2 }{b}\left(\cot \alpha_B+\int_{\alpha_B}^{\pi/2}\frac{r_s^2 /r_0^2}{(1-\frac{r_s}{r_0}\sin\alpha)}d\alpha \right)+\end{equation*}
\begin{equation*}+\frac{r_0 }{b}r_s \ln \left|\frac{1}{\sin\alpha_B}+\cot \alpha_B\right|\,.\end{equation*}
This expression is exact.

The integral in the parenthesis  can also be evaluated analytically but using that $r_s/r_0\ll 1$ and  $1-\frac{r_s}{r_0}\sin\alpha\geq 1-\frac{r_s}{r_0}$, it is significantly smaller then its counterpart $\cot \alpha_B$ when $ \alpha_B\ll 1$ and may be neglected. Thus, the time propagation between $P$ and $B$ is
\begin{equation}\label{shap}
c(T_B-T_P)\approx \frac{r_0 }{b}\left(x_B+r_s\ln \frac{r_B+x_B}{r_0}\right),
\end{equation}
where $x_B$ denote the $x$ coordinate of $B$. Using the above approximation and that $r_B\approx x_B$ the Shapiro time \textit{delay} for a signal traveling from $A$ to $B$ and back is
\begin{equation}\label{dTAB}
   r_s\ln \frac{4x_B|x_A|}{r_0^2},
\end{equation}
which is the known formula \cite{MTW}, \cite{Rindler}, \cite{GS} and \cite{KEK} for Shapiro time dilation. This formula  {has been confirmed} experimentally by several experiments, see \cite{CW}.

\section{Summary}

 In this paper we further refined  the RND model introduced earlier. The model treats gravity as a force and incorporates the influence of potential energy on spacetime in Newtonian dynamics, without  curving spacetime as in Einstein's GR. It is assumed that the \textit{potential} energy is the classical one defined by Newton's (linear) equation for the gravitational potential.

 For a gravitational potential $U(\mathbf{x})$ which is static in some inertial frame and vanishes at infinity, its influence in the neighbourhood of a space point is expressed by a local metric  (\ref{metric2}) preserving the symmetry of the influence. This metric is uniquely defined from $U(\mathbf{x})$ by assuming that this influence is the same as the influence of the velocity of the  escape frame on spacetime. The connection between time and the  affine parameter on the trajectory is defined via the conservation law using the Killing vector.

 The RND \textit{kinetic} energy is \textit{almost} the classical $\frac{mv^2}{2}$ with the difference that the norm of the velocity is altered due to the influence of the potential, and the  differentiation is with respect to the affine parameter on the trajectory instead of time. The RND energy equation (\ref{RNDHamit}) is now derived for both objects with non-zero as well as for \textit{massless} particles. The equations of motion with respect to the affine parameter (\ref{RND1finals}) and with respect to time (\ref{RNDf}) are derived from the conservation of the total energy on the trajectory. For a central force the RND model predicts also conservation of angular momentum.

 In our previous papers we have shown that the RND  passed successfully the first two tests of GR. In this paper we have shown that RND also predicts accurately the Shapiro time delay - the fourth test of GR. The gravitational deflection of any object/particle passing the strong gravitating field of a massive body will be treated in a forthcoming paper.

\acknowledgments {I wish to thank Prof. J. M. Steiner for the  discussions which helped to formulate and clarify the results of this paper, {the referees} and Dr. T. Scarr for their constructive comments.}

\end{document}